\documentclass{aa}
\usepackage{graphicx}
\begin{document}
\title
  {Photoionising feedback and the star formation rates in galaxies}
  \titlerunning{Photoionising feedback}
\author
  {J. M.~MacLachlan\inst{1},
 I. A.~Bonnell\inst{1},
 K.~Wood\inst{1},
 and J. E.~Dale\inst{2}}
\institute{SUPA, School of Physics \& Astronomy, University of St. Andrews, North Haugh, St. Andrews KY16 9SS
	\and
	Excellence Cluster 'Universe', Boltzmannstr. 2, D-85748 Garching, Germany}
 \authorrunning{MacLachlan et al.}
\date{Received 10 July 2013 / Accepted 11 October 2014}

\abstract
{}
{We investigate the effects of ionising photons on accretion and stellar mass growth in a young star forming region, using a Monte Carlo radiation transfer code coupled to a smoothed particle hydrodynamics (SPH) simulation.}
{We introduce the framework with which we correct stellar cluster masses for the effects of photoionising (PI) feedback and compare to the results of a full ionisation hydrodynamics code.}
{We present results of our simulations of star formation in the spiral arm of a disk galaxy, including the effects of photoionising radiation from high mass stars. We find that PI feedback reduces the total mass accreted onto stellar clusters by $\approx 23\%$ over the course of the simulation and reduces the number of high mass clusters, as well as the maximum mass attained by a stellar cluster. Mean star formation rates (SFRs) drop from SFR$_{control}=4.2\times10^{-2}$ M$_{\odot}$yr$^{-1}$ to SFR$_{MCPI}=3.2\times10^{-2}$ M$_{\odot}$yr$^{-1}$ after the inclusion of PI feedback with a final instantaneous SFR reduction of 62\%. The overall cluster mass distribution appears to be affected little by PI feedback.}
 {We compare our results to the observed extra-galactic Schmidt-Kennicutt relation and the observed properties of local star forming regions in the Milky Way and find that {internal photoionising (PI) feedback is unlikely to reduce star formation rates by more than a factor of $\approx 2$ and thus  may play only a minor role in regulating star formation}.}


\maketitle
\section {Introduction}
Feedback from young massive stars can have a significant impact on their environment on scales of tens of au to several kpc \citep{hollenbach_2000,franco_1994,tenorio-tagle_19792,whitworth_erosion_1979}, and is often invoked to regulate star formation rates in galaxies.  The emission of high frequency photoionising photons into the surrounding neutral gas can act to unbind the gas and disrupt ongoing star formation.
On small scales the action of PI photons from the central star is a principal mechanism for the dispersal of circumstellar disks around massive stars \citep{hollenbach_2000,bally_external_1998}. In dense cluster environments the photoevaporation by an external source can cause the destruction of circumstellar disks, even around low mass stars, if the cluster is large enough to host an O or B star. 

It is thought that a large fraction of stars form in clusters embedded within molecular clouds rather than in isolation \citep{lada_micron_1991,lada_cluster_2003} and so the action of massive stars can strongly affect the growth and evolution of other cluster members. In the case where ionised gas becomes unbound and escapes from the cluster,
then it can lead to the complete disruption of the stellar cluster \citep{hills_1980,Baumgardt:2007kg,Goodwin:2006fh}.

In addition to limiting the ability of stars to accrete neutral gas, the influence of ionised gas can also promote the formation of new stars. This may be due to the ``collect and collapse'' mechanism in which neutral gas is swept up by the advancing ionisation front and forms unstable clumps that undergo collapse and form stars \citep{whitworth_1994,elmegreen_1977,dale_2007}. Alternatively additional star formation may be caused by the radiatively driven implosion of pre-existing stable gas clumps that exist in the structure of the undisturbed molecular cloud \citep{Kessel_2003,bisbas_2011}.

It has been shown by previous authors \citep{dale_photoionizing_2005,dale_2011,krumholz_2005} that the underlying structure of the neutral gas in which an ionising source is embedded will have a strong influence on the propagation of ionising photons. If the source is accreting from high density accretion flows, then these will strongly inhibit the flow of PI photons in certain directions and effectively shield portions of the surrounding medium from the direct effects of the source. The ionising photons will escape more easily  into the low density cavities of the surrounding gas, and the accretion flows can remain mostly neutral. 

Smoothed particle hydrodynamics (SPH) provides a useful method to follow the evolution of a molecular cloud where large contrasts in gas densities are present.
The Lagrangian nature of SPH codes also helps in following the evolution of self-gravitating flows,  but traditional methods for evaluating the transfer of radiation in such environments are often best suited to Eulerian schemes, where the optical depth can be more simply integrated along directions of photon travel. Several authors have implemented radiation transfer methods in SPH simulations to treat PI radiation \citep{dale_photoionising_2005,dale_new_2007,gritschneder_ivine_2009,kessel_ionising_2000}. While methods exist to implement radiation transfer in SPH codes, many simulations still neglect such effects for the sake of computational speed and complexity.

We have used a Monte Carlo photoionisation (MCPI) code to calculate the ionisation structure for a series of static, independent snapshots of the SPH simulation. We do not re-run the SPH simulation but instead use the MCPI code to estimate which of the SPH particles become photoionised by young stars and remove their contribution to future star formation.
There is no additional computational overhead during the SPH simulation as all photoionisation calculations are performed once the simulation is completed. In this way we are able to characterise the reduction in SFRs and efficiency caused directly by PI feedback. The aim of this work is to provide an estimate of the first order effects of feedback from PI radiation. We first test our method against a full ionisation hydrodynamic simulation before applying it to star formation in a galactic spiral arm.

\section{SPH Simulations}
We begin our investigation with the star formation simulations of \citet{bonnell_2012}. These are designed to track the small scale star formation processes as well as the large scale evolution of the interstellar medium on galactic scales. Three simulations were carried out by \citet{bonnell_2012}. Firstly a global disk simulation of a $5-10$ kpc annulus of material within a galactic disk was evolved for $370$ Myrs. Cooling in the spiral shocks produced a complex, multiphase interstellar medium in which dense clouds reminiscent of the molecular clouds seen within our own galaxy formed. To follow the star formation process within such dense clouds a second high resolution run was then carried out, the \textit{Cloud} simulation. This followed the formation of one dense cloud at higher resolution than the initial global simulation. Finally, the \textit{Gravity} simulation added the effects of self-gravity and allowed the dense cloud to form stars from the realistic initial conditions provided by the \textit{Cloud} simulation. The \textit{Gravity} simulation contains $1.29\times10^7$ SPH particles with each particle having a mass of $0.15$ M$_{\odot}$. The formation of stars was treated with the use of sink particles \citep{bate_1995}. Due to the prescription of the formation of a sink particle which requires a minimum of $70$ particles we are only able to resolve the gravitational collapse to a mass of $\approx11$ M$_{\odot}$. This is too high to represent low mass star formation and each sink should instead be thought of as a stellar cluster.  The results from \citet{bonnell_2012} based on the \textit{Gravity} simulation suggest that the observed Schmnidt-Kennicutt\citep{schmidt_rate_1963,kennicutt_global_1998} power law relation between star formation rate and gas surface density can be explained by the combination of spiral shocks and gas cooling rates. However, the star formation rates produced by the simulation are higher than those observed. 

We start our investigation with the \textit{Gravity} simulation. The dense cloud it represents is approximately $250$ pc in size and contains $\approx 1.7\times10^{6}$ M$_{\odot}$ of gas. We assume that no prior star formation has occurred in the cloud or its surroundings during its previous evolution. The star formation and evolution of the cloud is traced for $\sim5$ Myrs. Beyond this point it is likely that supernovae within the cloud will quickly act to disperse it, ending most star formation. 
We assume that all gas forming or accreting onto a sink particle is retained by the sink and that none is recycled back into the simulation by feedback on the scale of the small clusters represented by the sinks.
Star formation efficiencies are unlikely to be this high and evidence suggests that they are in the region $25-50\%$ \citep{olmi_constraints_2002} which would lead to our star formation rates being upper limits by a factor $2-4$.


During the time evolution of the \textit{Gravity} simulation the properties of each SPH particle (position, velocity etc.) are printed out to file at several points. It is on these outputs that we base our analysis regarding the effects of photoionisation on the star formation process. 
\begin{figure}
\centering
\includegraphics[width=0.45\textwidth]{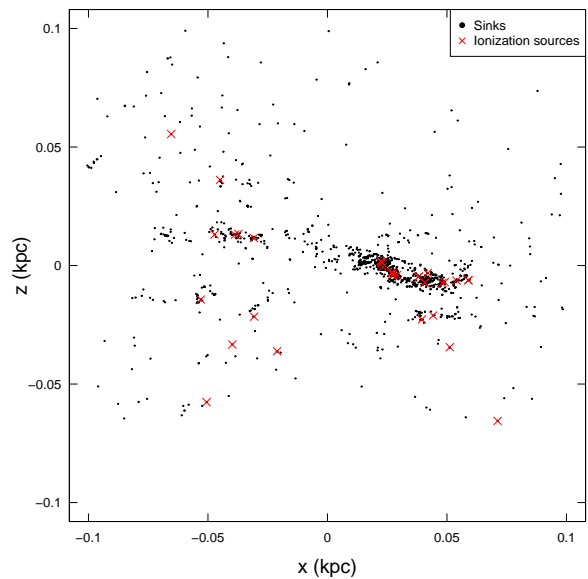}
\caption{Positions of the stellar sink particles (black circles) and the ionising sources at the end of the SPH simulation ($356$ Myrs) with $\Sigma_{gas} = 4$ M$_{\odot}$pc$^{-2}$. }
\end{figure}
\section{Monte Carlo Photoionisation Code}\label{mc}
In order to calculate the photoionisation of the gas throughout the star forming region during its evolution we use a MCPI code described in \citep{wood_escape_2000} and extended by \cite{wood_photoionisation_2010}. The routines can be used to calculate the hydrogen-only photoionisation and have been modified to allow the use of SPH data as initial input. The Monte Carlo treatment of the radiation transfer is fully 3D and can easily treat the complex geometries encountered within the gas distribution.

All ionising photons encounter an average opacity which is given by $n_{H_{0}}\bar{\sigma} + n_{H}\sigma_{dust}$, where $ n_{H_{0}} $ is the number density of neutral hydrogen, $\bar{\sigma}$ is the flux averaged cross section and $\sigma_{dust}$ gives the dust cross section per hydrogen atom (ie. cm$^{-2}$H$^{-1}$).
Two different flux averaged cross sections are considered depending on whether a photon has been emitted directly by an ionising source (direct photon) or is the product of absorption and subsequent reemission (diffuse photon). The cross section for direct photons is averaged over the flux \begin{equation}\bar{\sigma}=\frac{\int_{\nu_{0}}^\infty \! F\sigma_{\nu} \, \mathrm{d}\nu}{ \int \! F \, \mathrm{d}\nu},\end{equation} where $F$ is the source ionising spectrum, $\nu_{0}$ is the frequency of the Lyman edge and $\sigma$ is the absorption cross-section for Hydrogen. In the case of the diffuse ionising spectrum the photon energies are strongly peaked at just above $13.6$eV and so we set the cross section for the diffuse photons to the hydrogen cross-section just above $13.6$eV, $\bar{\sigma}=6.2\times 10^{-18}$cm$^{-2}$. This simplification means that essentially only two frequencies are considered and greatly speeds the computation of the ionisation fractions throughout the grid.

As an output of the photoionisation code we have a grid containing the neutral fraction ($f_{n}$) which gives the fraction of the mass in each grid cell which is neutral, with the remainder being ionised.
In order to apply our MCPI routines to the SPH data it is necessary to discretise the SPH density estimate across a cartesian grid. In this case each SPH simulation dump is discretised onto a $200^{3}$ grid of cells. For the  \textit{Gravity} simulation we centre the grid on the centre of mass of the system and have the grid extend to $\pm0.1$kpc in all three axes, to fully encompass the central region of the cloud with sufficient resolution. This results in a 1pc resolution for the MCPI simulations. { We  found that an increased resolution (a $2^3$ increase in grid cells)  has little impact on the results 
(of order a 1 per cent change in the ionised gas mass) of the MCPI code, but adds significantly to the time required for each simulation.}  

\subsection{ionising Photon Numbers}\label{ionflux}
To estimate the positions and locations of ionising sources within the molecular cloud we create a list of all stellar sink particles in the simulation along with their locations. Due to the limited mass resolution of the simulations each sink particle represents a stellar cluster in this case. The number of photons able to ionise hydrogen produced each second by each stellar cluster is represented by the $Q_{H}$ value. Following the method of \citet{dale_2011} we assign a $Q_{H}$ to clusters that exceed $600$ M$_{\odot}$ by calculating the total mass in stars 
M$>30$ M$_{\odot}$ and dividing this mass by $30$ M$_{\odot}$. This gives an approximate measure of the number of $30$ M$_{\odot}$ stars, N$_{30}$.
To estimate $Q_{H}$ for a $30$ M$_{\odot}$ O-star we fit the data of \citet{Conti_2008} (Table 3.1) and \citet{diaz-miller_1998} (Table 1) to find a relation between the stellar mass and number of ionising photons. From this we estimate that the number of ionising photons produced by a 
$30$ M$_{\odot}$ star will be $Q_{H}=2.8\times10^{48}$ s$^{-1}$. 

{We assume a Salpeter IMF  with upper and lower mass limits of $100$ M$_{\odot}$ and $0.1$ M$_{\odot}$. The use of a Salpeter IMF slope down to $0.1$ M$_{\odot}$ is  an overestimate of the total mass in low mass stars as the true IMF should turn over near $0.5$ M$_{\odot}$ \citep{kroupa_distribution_1993}. This will lead to lead to under-estimates for the mass in stars $>30$ M$_{\odot}$, but is likely offset by our assumption of as 100 per cent star formation efficiency. We performed a test simulation using a two-part Kroupa IMF \citep{kroupa_variation_2001} and included mean values of $Q_H$ from stochastic sampling. The ionising fluxes in this case was typically a factor of 2 higher than given by our regular estimate,  but this increase is less than that due to our assumed 100 per cent star formation efficiency, which likely overestimates the total stellar mass by a factor $2-4$. Dale et al. (2012a) find that uncertainties in the ionising fluxes of factors of a few have little impact in the ionised gas mass. We can reasonably conclude that  these simulations are likely to encompass  the maximum impact of photo-ionising  feedback.  }
It is possible that the stochastic sampling of the IMF could result in significantly larger fluxes at low cluster masses in rare cases with a high-mass star in isolation or in
a low-mass cluster. We do not consider such cases here.ionis

\section{Coupling the MCPI to SPH simulations}\label{coupling}
As the time evolution of the SPH simulation we are investigating has previously been completed and did not include the effects of photoionisation, it is necessary to approximate the feedback effects as well as possible without rerunning the simulation itself. By linking our MCPI codes to the individual time dumps of the SPH code we hope to investigate the effects that photoionisation by young stars can have on the mass accretion rates during the course of the simulation. To accomplish this we have included a series of checks on each SPH gas particle, that are carried out at each SPH time dump in the simulation. 

We initially begin at the point in the simulation where at least one stellar sink has a mass $\geq 600$ M$_{\odot}$, as this will provide the first source of ionising photons as described in Section \ref{mc}. We then run our MCPI code using as inputs the SPH density discretised into a cartesian grid and the positions and $Q_{H}$ values for the stellar clusters. The output of the the MCPI code is a grid containing the hydrogen ionisation fraction throughout the star forming region at the time-step of interest. 

Once the MCPI code has computed the ionisation fraction within the grid we identify the SPH particles that are located within ionised cells. If a gas phase SPH particle is located in a cell with an neutral fraction ($f_{n}$) $<0.5$ then it is classified as an ionised particle. We have found that in general the results we obtain are relatively insensitive to the exact value of $f_{n}$ at which particles are classed as ionised, as the mass of ionised particles is dominated by SPH particles that are almost fully ionised ($f_{n}<0.1$).

We adjust the accretion rates by removing any mass accreted from an ``ionised'' SPH particle as such gas should no longer be gravitationally bound to the accreting sink. The stellar cluster masses are recalculated by removing an amount of mass from the sink equal to the mass of the ionised particle multiplied by ($1-f_{n}$). In this way if an accreted ionised particle was found in a cell with $f_{n} = 0.4$ then $60\%$ of its mass would be prevented from accreting onto the stellar sink. { ionised gas is assumed
to escape the simulation and be no longer available for star formation. This means that the ionisation fraction cannot decrease. 
In testing this approach, we found that few particles ever returned to a neutral status as once they are exposed to the ionising radiation, they did not
re-enter a shadowed region in these simulations. }
We also test for the ionisation of the SPH particles that are used to initially form stellar sinks. If, when a sink particle is formed, more that $50\%$ of the gas particles that took part in its formation were classified as ionised, then the sink will be removed from further calculations and assumed not to form. 

This method allows us to investigate the effects of ionisation feedback on the accretion onto stellar clusters as the simulation progresses and compare the results with the initial distributions that do not include any effects of photoionisation by massive stars.

{It is important to note that no dynamical effects of the photoionisation are included. Thus the pressure from the hot gas onto the surrounding environment is ignored as is
the reduction of gravitational acceleration from the reduced sink masses in the simulation. The gravitational and pressure forces which were used to determine the dynamics 
are entirely due to the non-ionising runs with higher sink masses and fully neutral gas. This is an unfortunate but unavoidable effect of our approach and therefore we have
first compared this approach to the full dynamical ionisation simulations from Dale et al (2012). The fact that we are able to largely replicate the Dale  results (see below) has allowed us to proceed with this approach with a high level of confidence in the results. }

\section{Validity of the Method}
In order to assess how well the methodology we have adopted is able to estimate the effects of ionisation feedback on the growth of stellar mass in our simulations we have performed tests against the simulations of \citet{dale_2012} (hereafter referred to as D12). We have used the results from the run A and run I simulations from D12 in order to test the effectiveness of our treatment under differing conditions and their properties are summarised in table \ref{daleProps}. Run A has the largest radius of the simulations to form any stars in D12 and shows diffuse structure. It contains a few tens of clusters, with a number that are massive enough to become sources of ionising photons. Run I is physically much smaller and more compact than run A and possesses the lowest escape velocity of any of the simulations investigated. In run I the mass resolution is sufficient to consider the sink particles that form as individual stars and so we can directly estimate number of ionising photons produced from the stellar mass. We adopt the same method as D12 to calculate the $Q_{H}$ values in run I and assign any sink that exceeds $20$ M$_{\odot}$ a number of ionising photons according to the relation:
\begin{equation}
log(Q_{H}) = 48.1 + 0.02(M_{\star} - 20M_{\odot}),
\end{equation}
 where $M_{\star}$ is the stellar mass. In run A we adopt the standard relation between cluster mass and number of ionising photons from section \ref{ionflux}.
 
 For both run A and I we have access to the control simulations, which included no ionisation effects (hereby referred to as the control runs), and also the simulations including the effects of ionisation. We apply our methods to the control runs and compare the results to the simulations which explicitly included ionisation. In this way, we can assess how important the dynamics of the ionised gas are, which we cannot follow in our MCPI method.
\begin{table*}
\begin{center}
\caption{Properties of the simulations from D12 used to test the MCPI method.}
\begin{tabular}{l l l l l l l }
\hline
Run&Mass&Radius&$v_{rms}$&$\langle n(H_{2})\rangle$&t$_{ff}$&$\langle T \rangle$\\
&(M$_{\odot}$)&(pc)&(kms$^{-1}$)&(cm$^{-3}$)&(Myr)&(K)\\
\hline
A&$10^{6}$&$180$&$5.0$&$2.9$&$19.6$&$143$\\
I&$10^{4}$&$10$&$2.1$&$136$&$2.56$&$53$\\
\hline
\label{daleProps}
\end{tabular}
\end{center}
\end{table*}


%
For run A we have an excellent agreement between our calculations and D12 over most of the time covered by the simulation. The integrated SFR, calculated from the total stellar mass formed during the simulation is $SFR_{control}=3.6\times10^{-3}$ M$_{\odot}$yr$^{-1}$ in the control and drops to $SFR_{Dale}=2.8\times10^{-3}$ M$_{\odot}$yr$^{-1}$ and $SFR_{MCPI}=2.5\times10^{-3}$ M$_{\odot}$yr$^{-1}$ in the D12 and MCPI simulation respectively. 

For run I the integrated SFR for the control is $SFR_{control}=1.1\times10^{-4}$ M$_{\odot}$yr$^{-1}$ and drops to $SFR_{Dale}=7.1\times10^{-5}$ M$_{\odot}$yr$^{-1}$ for the D12 simulation and $SFR_{MCPI}=5.7\times10^{-5}$ M$_{\odot}$yr$^{-1}$in the MCPI run.

\subsection{Cumulative Stellar Mass}
\begin{figure}
\centering
\includegraphics[width=0.45\textwidth]{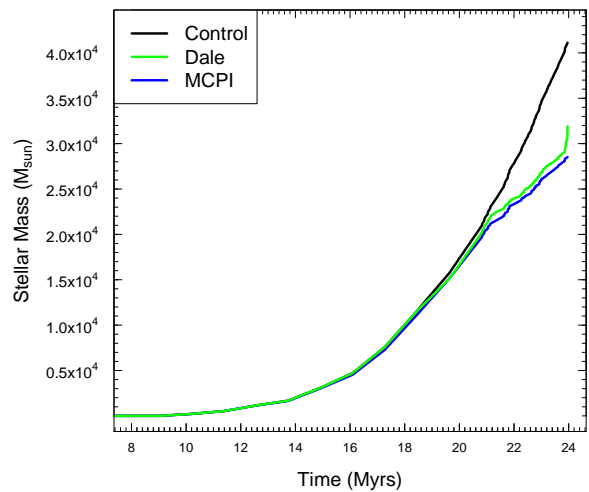}
\caption{Cumulative stellar sink mass formed during run A for the  control (black), Dale photoionisation (green) and MCPI (blue) runs. }
\label{CSRunA}
\end{figure}

\begin{figure}
\centering
\includegraphics[width=0.45\textwidth]{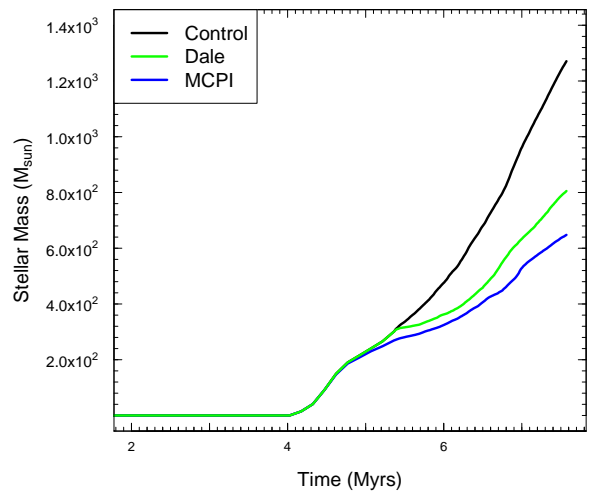}
\caption{Same as figure \ref{CSRunA} but for run I.} 
\label{CSRunI}
\end{figure}

In figures \ref{CSRunA} and \ref{CSRunI} we present the cumulative stellar sink mass formed over the course of the run A and run I simulations in the control, D12 and MCPI implementations. In run A we have a good agreement between the two methods at most times and the difference in the stellar mass is less than $5\%$ for the majority of the simulation, with the exception of the final few timesteps where the SFR in D12 increases significantly with a corresponding change in the stellar mass.

The results from run I are also in reasonable agreement. The MCPI tends to overestimate the effects of feedback at all times with a steady divergence of the stellar sink masses over the course of the simulation. By the end of the run the difference is around $20\%$. Again the difficulties of the MCPI method in correctly treating run I are not unexpected due to the strong dynamical influence of the ionised gas.

The fact that the MCPI results are in reasonable agreement with the more sophisticated D12 treatment gives us encouragement that we are able to treat the ionisation feedback on stellar growth in a satisfactory way. It should also be noted that mostly the MCPI method predicts a SFR that is equal to or lower than D12. The MCPI treatment can then be interpreted as providing an upper limit to the effects of ionising feedback by only allowing it to reduce the SFRs and not accounting for the possible triggering of star formation.

\section{Results}
\begin{figure}
\centering
\includegraphics[width=0.45\textwidth]{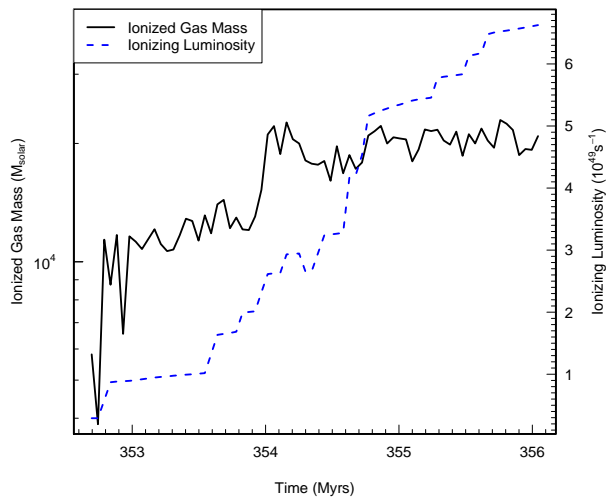}
\caption{{Evolution of the ionised gas-mass and ionising luminosity  are plotted as a function of time from the MCPI simulation. The ionised gas mass shown here does not include any gas that has been ionised previously, but was accreted in our control run at the same point in time. Thus the total ionised mass implied by our MCPI simulations is the sum of the ionised gas
present at each timestep (shown here) and that subtracted from the stellar mass due to being ionised (Figure~\ref{cumulative}).}}
\label{time}
\end{figure}

\begin{figure}
\centering
\includegraphics[width=0.45\textwidth]{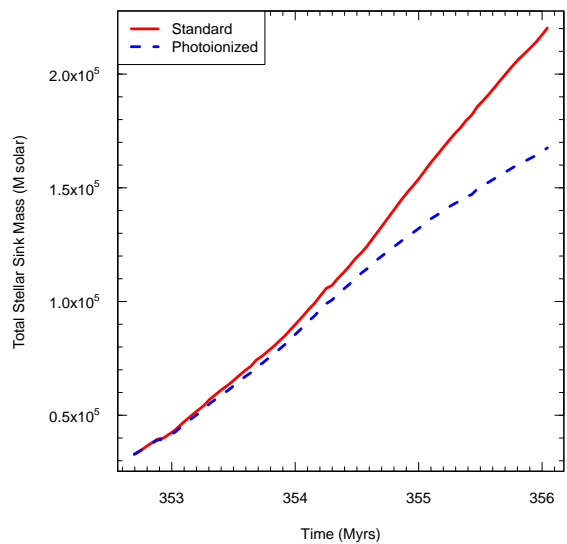}
\caption{Total stellar mass is plotted as a function of time for the standard (red-solid) and photoionised (blue-dashed) simulation. The difference represents the mass of gas that is removed from the star formation process due to being photoionised.}
\label{cumulative}
\end{figure}

In Figure \ref{time} we plot the time evolution of the ionised gas mass and ionising luminosity as calculated by the MCPI simulation at each timestep from our SPH simulation. The total ionising luminosity grows steadily throughout the simulation from $0.29 - 6.6 \times10^{49}$s$^{-1}$as the sink particles accrete more mas. Each step in the ionising luminosity represents a new sink particle (or particles) exceeding the minimum mass required to become an ionising source. The decrease in ionising luminosity at $\sim 354.3$Myrs is due to an ionising source exiting the region of interest. In our work the ionised gas mass is representative of the total number of SPH gas particles, present in each timestep, which have been ionised and therefore excluded from any subsequent star formation. The ionised gas mass shows an initial fast growth as the first ionising sources appear, followed by a modest, steady growth as the simulation progresses. As each MCPI simulation is independent, the ionised gas mass can decrease in time due to the exact distribution of sources and neutral gas in each case. {It can also decrease if any ionised gas is subsequently accreted in our control runs, and therefore not present to be ionised. Thus, the total amount of ionised gas implied in our simulations is not just the gas ionised at each time step (Figure \ref{time}), but also includes any accreted gas that was previously ionised and therefore subtracted from the total stellar mass as shown in Figure~\ref{cumulative}. }

\begin{figure*}
\centering
\includegraphics[width=0.98\textwidth]{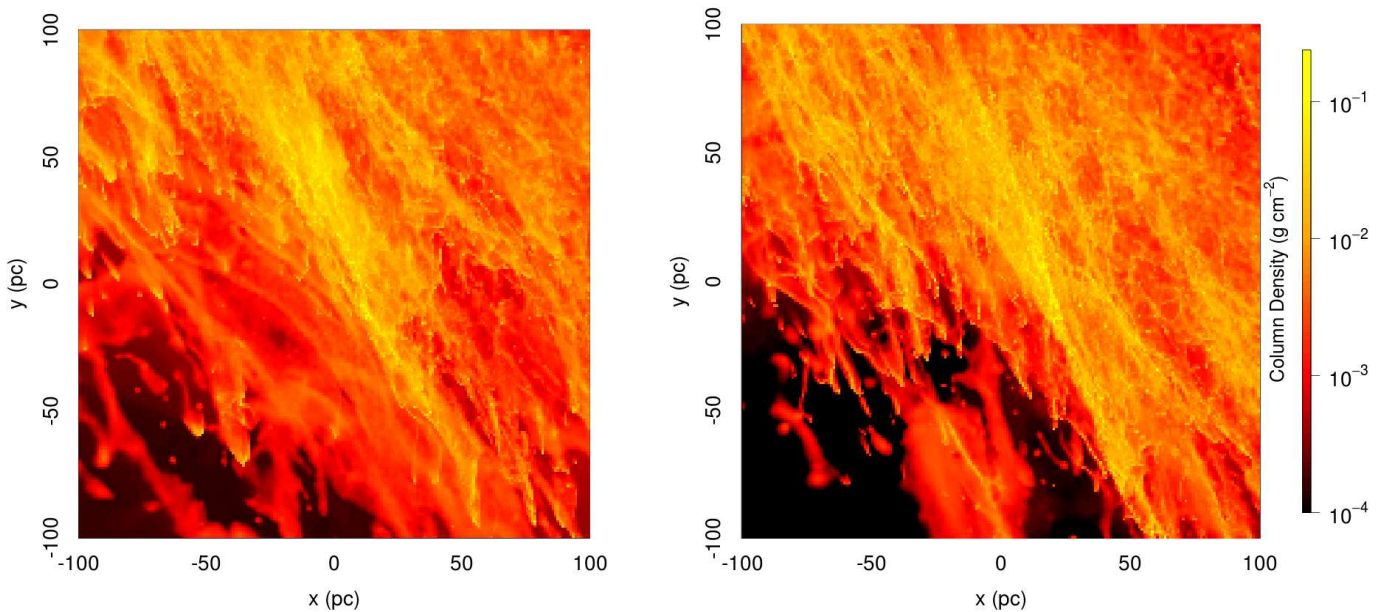}
\caption{Distribution of the neutral gas at the start (left) and end of the MCPI simulation (right).}
\label{neutralgas}
\end{figure*}

\begin{figure*}
\centering
\includegraphics[width=0.98\textwidth]{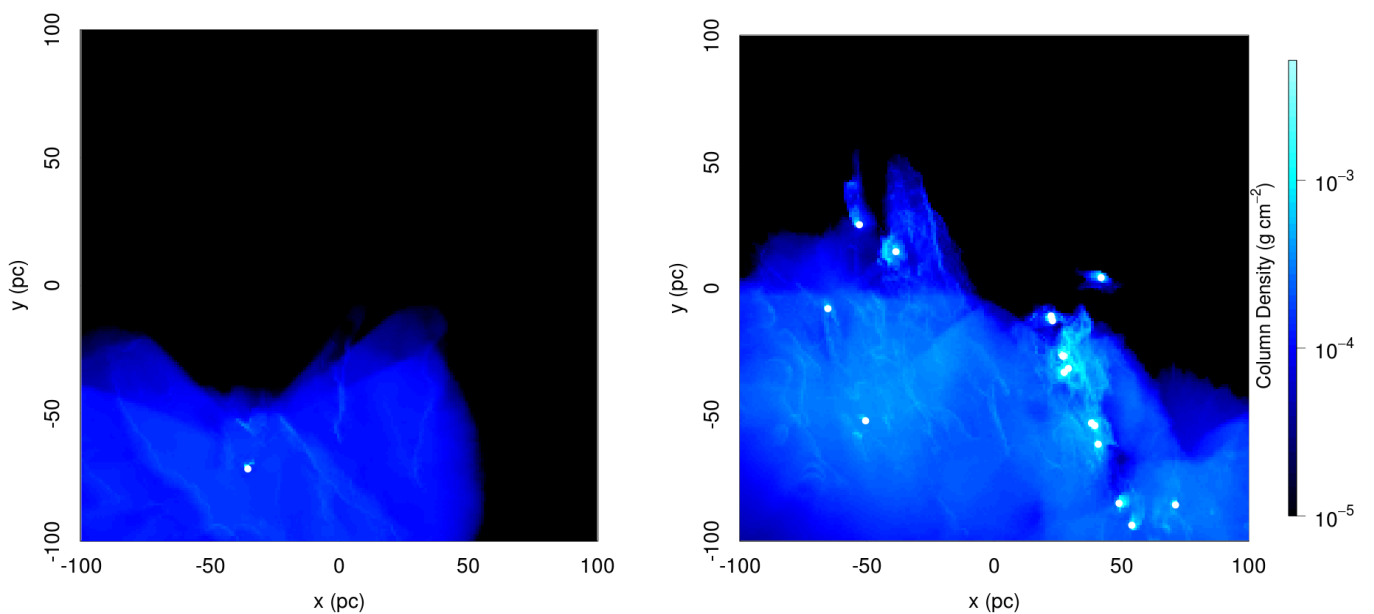}
\caption{Distribution of the ionised gas particles at the start (left) and end of the MCPI simulation (right). Brighter regions have a higher fraction of the gas ionised and hence removed from the star formation process. The positions of the ionising sources are marked by filled white circles.}
\label{iongas}
\end{figure*}

Figure \ref{iongas} shows the distribution of the ionised gas particles at the first time step containing an ionising source (left panel) and at the end of the \textit{Gravity} simulation. Figure \ref{neutralgas} shows the neutral gas surface density at the corresponding times. It can be seen that even in the early stages of PI feedback the ionising photons are able to affect a large region of the cloud. However, much of the gas which has been ionised at large distances from the ionising sources is low density, diffuse gas, that is not currently actively star forming. By the end of the \textit{Gravity} simulation the ionised gas has filled a large region of the cloud, with highly ionised regions within the dense inner core of the cloud close to the most massive ionising sources. The region of the cloud in the upper section of the plot ($y>0$)  is largely unaffected by the PI photons. 
{This is due to a reduced star formation rate in this region and in the shielding from the high density gas in the central region of the cloud, preventing the propagation of ionising photons in this direction. This has the effect that the low-level of star formation in the upper region ($y>0$) is largely unaffected by PI feedback from the massive clusters near the cloud centre.}

%
%

In figure \ref{cumulative} we show the cumulative mass which has been accreted onto the stellar sink particles as a function of time from the point at which the first ionising source forms until the end of the simulation. {It can be seen that there is a steady divergence of the standard and photoionised cases due to the ionisation of gas particles, any accretion of which is discounted 
in our photoionisation models. We stress that this is simply a reduction in the estimate of the mass in sinks and the corresponding star formation rates. The actual dynamics are
unaffected as they were calculated in our control, non-ionising SPH simulation.}  We find that, in total, after our inclusion of photoionisation effects the final mass accreted by all sinks falls from M$_{control} = 2.2\times10^{5}$ M$_{\odot}$ to M$_{MCPI} = 1.7\times10^{5}$ M$_{\odot}$ which is a reduction in the stellar mass of $\approx23\%$.

We estimate the mean SFR over the course of the simulation for both runs by taking the total mass formed by the end of the simulation and dividing by the total time for the \textit{Gravity} simulation. In the control run  we achieve SFR$_{control} = 0.042$ M$_{\odot}$yr$^{-1}$, while in the case including photoionisation the rate drops to SFR$_{MCPI} = 0.032$ M$_{\odot}$yr$^{-1}$. This change is due to feedback from ionising sources which have previously formed, ionising gas before it can be accreted onto stellar sinks or from new sinks.

\begin{figure}
\centering
\includegraphics[width=0.45\textwidth]{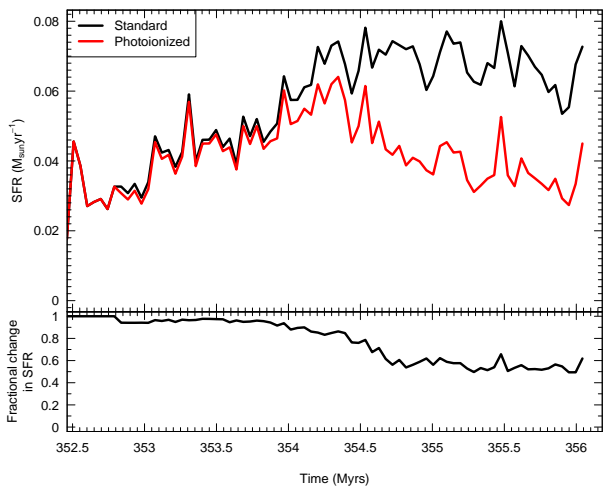}
\caption{Star formation rates of the control and MCPI simulations as a function of time.} 
\label{SFR}
\end{figure}

The effect of the PI feedback on the SFR appears to increase as the simulation progresses. The upper panel in figure \ref{SFR} shows the calculated SFR for the control and MCPI simulations and the lower panel shows the fractional difference between the two. Initially, before the formation of any ionising sources, both have identical SFRs. There is an immediate drop in the MCPI SFR compared to the control run as soon as the first massive clusters form followed by a period where the reduction in SFR due to PI feedback is around $5\%$. The last $\sim2$Myrs show a steady and generally gradual increase in the effects of PI feedback on the SFR. By the end of the simulation the SFR in the MCPI run is $62\%$ of that in the control run.
 
\begin{figure}
\centering
\includegraphics[width=0.45\textwidth]{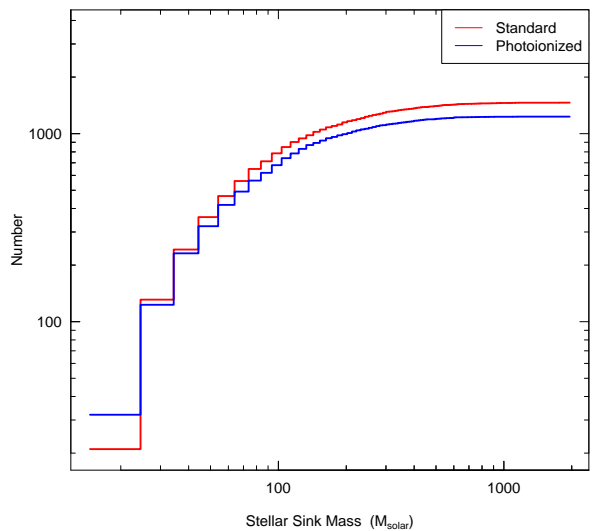}
\caption{Final cumulative stellar mass distributions for the standard (red) and photoionised (blue) runs.}
\label{hist}
\end{figure}

Figure \ref{hist} shows a cumulative histogram of the final stellar sink masses for both the standard and photoionised simulations. It shows the total numbers of sinks formed by the end of each simulation as a function of their mass. It can be seen that fewer sinks have formed in the photoionised case compared to the standard case but both follow a similar distribution. The most obvious change in the sink mass distribution takes place at the high mass end where there are a lower number of sinks due to the action of PI feedback. The effects of photoionisation are to reduce the accretion rates and also the maximum mass attained by a sink during the evolution of the system. In the standard case sinks are allowed to grow freely, accreting gas without the effect of feedback and in this case we end up with several sinks with masses $> 1000$ M$_{\odot}$ and the highest sink mass of $ 1968$ M$_{\odot}$. In the photoionised simulation the highest mass sink is $1154$ M$_{\odot}$ and fewer of the sinks grow in excess of $600$ M$_{\odot}$. Here the effects of feedback have prevented the sinks accreting gas as it becomes ionised before its accretion and effectively reduces the source of gas available to the high mass clusters. 

\subsection{Additional Simulations}

{In addition to the case outlined above we have also investigated the effects of PI feedback on two analogous simulations but with mean surface densities ten times larger ($40$ M$_{\odot}$pc$^{-2}$) and ten times smaller ($0.4$ M$_{\odot}$pc$^{-2}$ than the standard case. These simulations were produced by adapting the SPH particle masses and then rerunning the simulations through a 50 Myr
evolution before introducing self-gravity. This ensures that the thermal physics of the gas, and the cloud formation process, is done self-consistently (see \citealt{bonnell_2012} for more details). In the
absence of photoionising feedback, the three simulations combine to reproduce a Schmidt-Kennicutt star formation relation but at a higher level than is observed.}
The low surface density simulation suffers from a lack of star formation due to a more limited reservoir of gas available for accretion and a lack of dense clumps where the gas can cool efficiently \citep{bonnell_2012}. This results in a much reduced star formation rate and a lack of high mass sinks within the SPH simulation. With no sinks greater in mass than $\approx 75$ M$_{\odot}$ we have no sources of PI photons as defined in section \ref{mc} and so are unable to follow the possible effects upon the star formation process.

\begin{figure}
\centering
\includegraphics[width=0.45\textwidth]{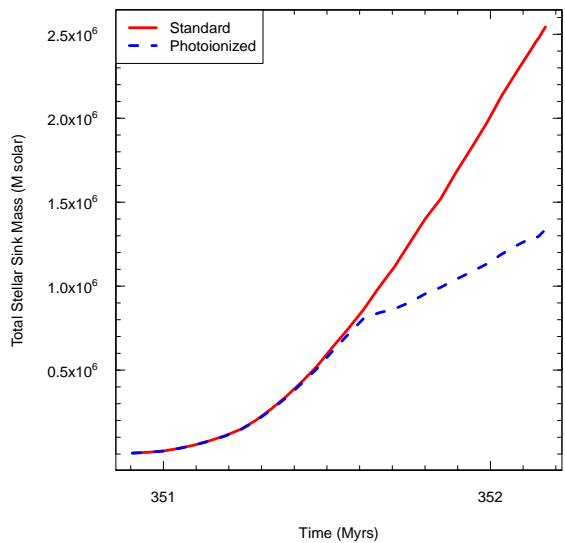}
\caption{Cumulative stellar mass as a function of time for the standard (red-solid) and photoionised (blue-dashed) simulation for the SPH simulation with a mean surface density of $40$ M$_{\odot}$pc$^{-2}$.}
\label{cumulativeHM}
\end{figure}

The high surface density SPH simulation ran for a shortened period of time due to the high densities producing prohibitively short time-steps within the SPH code. However, a large number of sink particles were able to form and grow in mass until they became sources of PI photons. A total stellar mass of $2.54\times10^{6}$ M$_{\odot}$ forms in the control run within the $\approx 1.3$Myrs which the simulation was evolved for and this is reduced to $1.33\times10^{6}$ M$_{\odot}$ in the MCPI simulation. Even in this short timescale the PI photons are able to strongly affect the star formation in the region. Figure \ref{cumulativeHM} shows the cumulative stellar mass in this case for the standard and photoionised cases. The mean star formation rate over the simulation falls from SFR$_{control} = 1.94$ M$_{\odot}$yr$^{-1}$ to SFR$_{MCPI} = 1.01$ M$_{\odot}$yr$^{-1}$. The sharp decrease in the star formation rate at a time of $\approx 351.6$Myrs appear to be caused by the coalescence of the ionised regions around several sources within the centre of the cloud to form one large ionised complex. This effectively ionises the majority of the gas near the sinks and removes the reservoir of star forming gas. 

\subsection{SFRs vs Surface Density}
In trying to understand the effects of PI feedback on the accretion processes it is useful to compare our results with the observed properties of star forming regions. It has been shown that there is a strong observed link between star formation and the local cold gas density \citep{schmidt_rate_1963,kennicutt_global_1998,kennicutt_2012}. In the case of disk averaged properties of local galaxies this can be fitted by a power-law relation of the form \begin{equation} \Sigma_{SFR} = A\Sigma_{g}^{N}, \end{equation} where $\Sigma_{SFR}$ is the star formation surface density, $\Sigma_{g}$ the total cold gas mass density and the power-law slope $N$ has a value of $\approx 1.4$ \citep{kennicutt_global_1998}. Observations of molecular clouds within our local environment in the Milky Way have suggested that the star formation surface density is a factor $10-20$ higher than that predicted by the extragalactic relation \citep{evans_spitzer_2009,heiderman_star_2010}. 

\begin{figure*}
\centering
\includegraphics[width=0.95\textwidth]{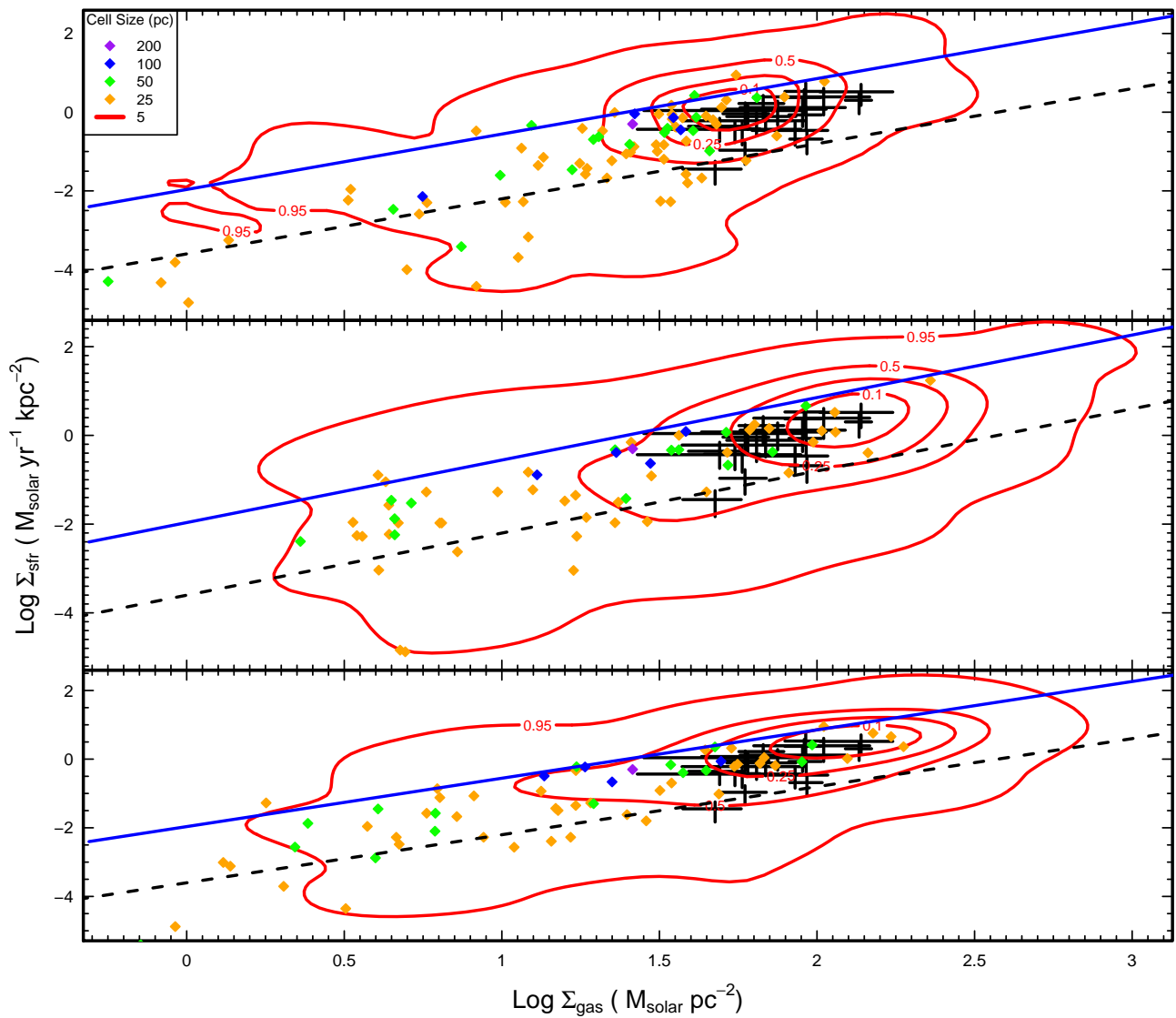}
\caption{Surface densities of the star formation rates are plotted as a function of the surface densities of the neutral gas.  The three panels represent the simulation as viewed along the three principle axes of the simulation grids. Crosses are observed data from Table 1. of \citet{heiderman_star_2010} with the extent of the cross indicting the confidence interval of the data. Coloured points (see key in top left) show the effects of increasing the cell size up to the size of the entire region ($200$pc). The dashed line shows the observed extragalactic power-law relation from \citep{kennicutt_global_1998} and the blue line the power law fit achieved by \citet{bonnell_2012} to the \textit{Gravity} simulation without PI feedback.}
\label{sfr2}
\end{figure*}

\citet{bonnell_2012} found that in the original SPH simulations there was a relation between $\Sigma_{SFR}$ and $\Sigma_{g}$ with a power law index $N=1.55 \pm 0.06$ (when the low ($0.4$ M$_{\odot}$pc$^{-2}$) and high ($40$ M$_{\odot}$pc$^{-2}$ surface density data is also included). The authors argue that the relation is a natural consequence of the shock formation of the dense, cold gas when cooling in the interstellar medium is included. In figure \ref{sfr2} we show the final instantaneous SFR and neutral gas mass surface density in our photoionisation simulation, as viewed along the three axes of our simulation grid. The contour lines and points show the effects of calculating the relationship within differing sized cells, from $5$pc to an average over the entire region. The values were calculated using grids of neutral gas density and SFR. Each grid was a $200^3$ cube of cells which we re-gridded to the required cell size before calculating the observed surface densities, as viewed along the three axes of the grid. In this way the entire cloud is covered at each resolution. The red contours show the data averaged over $5$pc cells with each contour enclosing a certain fraction of the data points, from $0.1$ to $0.95$. Only cells which contain a non-zero SFR have been plotted.  
We also show the observations of \citet{heiderman_star_2010} of local molecular clouds and YSO derived star formation rates, as well as the extra-galactic relation of \citet{kennicutt_global_1998}. Firstly it can be seen by comparing the three figures that the orientation of the cloud can influence its observed properties. The center and lower panels show a similar distribution of properties when observed at small scales (red contours) due to the fact they are both observed from within the mid-plane. The upper panel shows an offset to lower neutral gas surface densities and this is due its being observed from above the galactic plane where perspective effects reduce the surface density observed for different regions. As we increase the cell size over which the properties are averaged we see that, in general, all three figure move into better agreement. We see many regions in our cloud which overlap with the observed data of \citet{heiderman_star_2010} as well as many which seem to lie at lower gas surface densities.


%
%

\section{Discussion}
From the results shown in Figure \ref{cumulative} it is clear that the total mass accreted onto sinks throughout the simulation is overestimated by standard SPH methods when PI feedback is ignored. The final total stellar sink mass in the standard surface density simulation including photoionisation is reduced by around $23$\% and the instantaneous SFR is reduced to $62\%$ of its original value at the end of the simulation. This is a significant change and suggests that it is likely very important to include photoionisation effects to accurately model the SFRs and efficiency within SPH simulations.  
\subsection{Limitations}
The techniques outlined here provide a way to provide a first order approximation to the feedback provided by high mass stars. We neglect the dynamical effects of ionised gas that may provide additional forms of feedback due to the restrictions of our simulation.
\subsubsection{Triggering of additional star formation}
It has been suggested that the feedback from massive stars may act to trigger additional star formation under certain conditions. The expansion of the ionised HII region into surrounding neutral gas can sweep up a dense shell of material which can become self-gravitating \citep{whitworth_1994,elmegreen_1977,dale_2007}. Observationally it has proved difficult to definitively find evidence for triggered star formation surrounding massive stars but several authors have presented compelling arguments based on multi-wavelength datasets \citep{deharveng_2005,dirienzo_2012}. The extent to which these triggered star formation processes may counteract the reduction in star formation caused by the heating and ionisation of neutral gas is also unclear but simulations suggest they are likely second order effects \citep{dale_2012,Dale:2013da}.

There are also additional mechanisms in which the dynamical effects of ionised gas can allow the PI feedback to have a greater effect than we estimate here. The penetration of the ionising photons into dense regions of gas can be increased due to the streaming of hot ionised gas out of the cloud. \citet{whitworth_erosion_1979} and \cite{gendelev_2011} explored this scenario and found that the removal of ionised gas allows the ionisation front to penetrate deeper into dense gas and may increase the effects of ionising photons beyond those discussed here.

In addition to assuming a $100\%$ efficient star formation process the SPH simulation we have studied also neglect several physical processes which are likely to affect the overall star formation process. There is no treatment of magnetic fields which may cause a reduction in the star formation rate \citep{price_effect_2008,arthur_radiation_2011,mckee_1999} by providing an additional support mechanism to overcome self-gravity and prevent the collapse of less massive clouds to form stars. Self gravity is also not included in the initial disk simulation from which our dense star forming region originates. This will lead to a reduced kinetic energy which, again, will enhance star formation rates. The reduction in the star formation rates caused by the inclusion of such effects may provide a mechanism to move our simulated values into better agreement with observed relations.

We include no source of ionising photons outwith the bounds of our star forming cloud. In reality there will be a non zero background of ionising flux produced by previous bouts of star formation in the galactic disk. The effects of  background FUV heating are included in the SPH code \citep{VazquezSemadeni:2006je} but without account for the ionisation of the gas. \citet{reynolds_power_1990} found that the diffuse ionised gas (DIG) in our local environment requires a diffuse ionising flux of $\approx 4\times10^{-5} ergs~ s^{-1} cm^{-2}$ to maintain it. This additional source of PI photons may be able to affect the outer regions of the cloud, but is unlikely to penetrate into the dense cloud centre. While this is unlikely to affect the star formation in the cloud interior it may impact the initial stages of cloud formation during compression within the spiral shock.

\subsection{Localisation of stellar signatures}
We find that the clumpy gas distribution produced by the SPH calculation produces a highly structured complex of HII regions which, due to low density paths, provides an illustration of a possible source of ionising photons required to maintain the DIG seen in the Milky Way and external galaxies. At the final time-step of the simulation our photoionisation code suggests that $\approx 15\%$ of ionising photons escape into the ISM from this star forming complex. This escape fraction may also be important for observations of H$\alpha$ as a localised star formation indicator. If a significant fraction of the ionising photons are able to escape the star forming cloud then it also suggests there would be a corresponding drop in the H$\alpha$ luminosity. This effect will lead to H$\alpha$ emission that is non-coincident with the star formation. In turn this may cause deviations from the observed power law relation between star formation and gas surface densities at small scales \citep{relano_2012}. Indeed \citet{schruba_2010} and \citet{onodera_2010} have found observational evidence of such deviations in high spatial resolution studies of the star formation in the nearby spiral galaxy M33. Based on analysis of CO emission maps and H$\alpha$ imaging \citet{schruba_2010} estimate that the star formation law observed at large scales breaks down at scales below $\sim 300$pc. It is not possible to verify this effect here due to the size of the region studied.

In this work we have utilized a simple hydrogen only MCPI code but the method we propose can easily be extended to use radiation transfer codes which provide a more realistic treatment of the photoionisation physics \citep{2004MNRAS.348.1337W,ercolano_xray_2008}. Our simplified treatment likely only provides a first order correction as it focuses on the ionisation of hydrogen and the impact of Lyman continuum photons only, neglecting the effects of Helium as a source of opacity. There is, however, no reason that more complex radiation transfer codes which includes these effects could not be added to provide more accurate gas physics at the cost of computation time. The possibility to also self-consistently include calculations of star formation indicators, such as H$\alpha$ and $24\mu$m dust emission would allow the direct comparison with observational data. 

\section{Summary}
We have carried out an investigation into the effects of PI feedback from massive stars on the growth of stellar clusters in SPH simulations of a dense star forming region within a disk galaxy by utilizing post-completion MCPI calculations. In order to quantify the effects of PI photons we track SPH particles which are likely to be ionised during the course of the simulation and remove them from stellar sink particles where they have been accreted during the future evolution of the system. 

Results indicate that the total stellar mass in sink particles may be reduced by up to $23\%$ due to the action of PI photons emitted form the most massive stars. The final instantaneous SFR is reduced to $62\%$ of its original value.The overall distribution of sink particle masses does not seem to be greatly affected with the most noticeable changes occurring at the high mass end. The maximum sink mass attained is reduced and fewer massive stellar clusters form. As we cannot include the possible dynamical feedback effects of the hot ionised gas to trigger additional star formation our results provide an estimate of the maximum impact of PI feedback. 

A comparison of our region-averaged properties to observed extra-galactic relations and observations of local molecular clouds suggests that our simulations over-predict the star formation rate by a factor $2-10$. While the inclusion of PI feedback from massive stars is able to reduce the discrepancy between the simulation and observations it cannot fully account for it. We also find that a significant fraction of the PI photons emitted by massive stars in the simulation are able to traverse low density paths and escape the star formation region entirely. This provides a mechanism to produce a fraction of the DIG observed in many galaxies.
\section{Acknowledgements}
IAB acknowledges funding from the European Research Council for the FP7 ERC advanced grant project ECOGAL. This research was also supported by the DFG cluster of excellence `Origin and Structure of the Universe' (JED).
This research has made use of NASA's Astrophysics Data System Service.

\bibliographystyle{aa}
\bibliography{photoion}
\end{document}